\documentstyle[12pt,epsf]{article}
\renewcommand{\mathrm}[1]{{\rm #1}}

\begin{document}
\baselineskip 19pt
\parskip 7pt
\thispagestyle{empty}

\hfill  \today

\vspace{24pt}

\begin{center}
{\large\bf Two-level system with noise: Blue's function approach.
\vskip 0.5cm
}

\vspace{24pt}

\begin{center} 
	Ewa {\sc Gudowska-Nowak}\ ${}^{1}$
	G\'abor {\sc Papp}\ ${}^{2}$
	and J\"urgen {\sc Brickmann}\ ${}^{3}$
\end{center}

\vspace{8pt}
${ }^{1}$ {\sl Department of Physics,
Jagiellonian University,} \\
{\sl ul. Reymonta 4, 30-059 Krak{\'o}w, Poland;} \\
${ }^{2}$ {\sl GSI, Plankstr. 1, D-64291 Darmstadt, Germany \&}  \\
	{\sl Institute for Theoretical Physics, E{\"o}tv{\"o}s University, 
	H-1088 Budapest, Hungary;}\\
${ }^{3}$  {\sl Institute for Physical Chemistry
Technische Universit\"at Darmstadt,} \\ 
{\sl Petersenstr. 20,
D-64287, Germany;} 
\date\today 

\end{center}

\vspace{6pt}

\begin{center}
{\bf Abstract}
\end{center}
By using the random matrix approach and generalized Blue's function
representation  we solve analytically 
the model of an effective two-level system coupled to a noise reservoir.
We show that calculated spectral properties of the system  are in
agreement with  the numerically simulated results. We outline possible
applications of the model in the field of condensed phase reactions.
\newpage
\addtocounter{page}{-1}

\newcommand{\gm}{\gamma}
\newcommand{\ee}{\epsilon}
\renewcommand{\th}{\theta}
\newcommand{\Sg}{\Sigma}
\newcommand{\dl}{\delta}
\newcommand{\SSg}{\tilde{\Sigma}}
\newcommand{\eq}{\begin{equation}}
\newcommand{\eqx}{\end{equation}}
\newcommand{\eqn}{\begin{eqnarray}}
\newcommand{\eqnx}{\end{eqnarray}}
\newcommand{\ben}{\begin{eqnarray}}
\newcommand{\een}{\end{eqnarray}}
\newcommand{\f}[2]{\frac{#1}{#2}}
\newcommand{\ra}{\rangle}
\newcommand{\la}{\langle}
\newcommand{\bra}[1]{\la #1|}
\newcommand{\ket}[1]{| #1\ra}
\newcommand{\GG}{{\cal G}}
\renewcommand{\AA}{{\cal A}}
\newcommand{\GR}{G(z)}
\newcommand{\MM}{{\cal M}}
\newcommand{\BB}{{\cal B}}
\newcommand{\ZZ}{{\cal Z}}
\newcommand{\DD}{{\cal D}}
\newcommand{\HH}{{\cal H}}
\newcommand{\RR}{{\cal R}}
\newcommand{\arr}[4]{
\left(\begin{array}{cc}
#1&#2\\
#3&#4
\end{array}\right)
}
\newcommand{\arrd}[3]{
\left(\begin{array}{ccc}
#1&0&0\\
0&#2&0\\
0&0&#3
\end{array}\right)
}
\newcommand{\tr}{\mbox{\rm tr}\,}
\newcommand{\One}{\mbox{\bf 1}}
\newcommand{\pauli}{\sg_2}
\newcommand{\cor}[1]{<{#1}>}

\newcommand{\br}[1]{\overline{#1}}
\newcommand{\phib}{\br{\phi}}
\newcommand{\psib}{\br{\psi}}
\newcommand{\zb}{\br{z}}
\newcommand{\qb}{\br{q}}
\newcommand{\lm}{\lambda}
\newcommand{\ksi}{\xi}

\newcommand{\Gb}{\br{G}}
\newcommand{\Vb}{\br{V}}
\newcommand{\Gm}{G_{q\br{q}}}
\newcommand{\Vm}{V_{q\br{q}}}

\newcommand{\ggd}[2]{\GG_{#1}\otimes\GG^T_{#2}\Gamma}

\section{Introduction}

The theory of open quantum systems has been found to play a key role in
many areas of theoretical physics~\cite{WEISS}. The growing interest
focuses on quantum dynamics in condensed phase systems, to which the
most widely used approach is the reduced-density-matrix
theory~\cite{BLOCH,REDFIELD,HAAKE,MAHAUX}. While a
microscopic treatment of dissipative motion involves hermitean
Hamiltonians, use of non-hermitean Hamiltonians serves as a convenient
way to give a reduced description where the hermitean part of the
Hamiltonian refers to free undamped dynamics and the remainder describes
a damping imposed on the system by some external noise source ("heat bath").

\noindent Statistical properties of complex systems can be successfully
investigated within the framework of the random-matrix
theory~\cite{MEHTA} which turned out to be quite general and a powerful
phenomenological approach to a description of various phenomena such as
quantum chaos~\cite{GUTZWILLER}, complex nuclei~\cite{PORTER}, chaotic
scattering~\cite{MAHAUX} and mesoscopic physics~\cite{HAAKE}. Aspects of
vastly different physical situations such as electron localization
phenomena in disordered conductors and semiconductors~\cite{IIDA},
disordered quantum wires~\cite{MIRLIN} and quantum  Hall effect
\cite{WEIDE} can be described in the language of the random matrix
theory. In  all the realms mentioned above, the Hamiltonian of the
system is rather intricate to be handled or simply unknown. In such
cases the integration of the exact equations is replaced by the study of
the joint distribution function of the matrix elements of the
Hamiltonian $P({\bf H})$. If there is no preferential basis in the
space of matrix elements, so that the system is as random as possible
and equal weights  are given to all kinds of interactions, one has to
require the probability $P({\bf H})d{\bf H}$ to be invariant under
similarity transformations ${\bf H}\rightarrow R^{-1} {\bf H} R$ with
$R$ being orthogonal, unitary or a symplectic $N\times N$ matrix
reflecting the fundamental symmetry of the underlying Hamiltonian.

\noindent The natural way of addressing the problems of randomness
coupled to various sources is to use technique of the ``free random
variables''~\cite{VOICULESCU,ZEE,BREZIN,JANIK}. This method provides an
elegant way of ``linearizing'' the process of determining the average
eigenvalue distributions for convolutions representing  an analogue of
the logarithm of Fourier transformation of the  usual convolutions.
Recently, the generalization  of the ``addition law'' for hermitean
random matrices to the non-hermitean case has been derived~\cite{JANIK}.

\noindent In section~2, we introduce, after {\it Zee}~\cite{ZEE} and
{\it Janik et al.}~\cite{JANIK} basic concepts of the ``addition laws'', 
for both hermitean and non-hermitean ensembles.

\noindent In section~3, we apply the addition law to the dissipative system 
of a two-level deterministic Hamiltonian coupled to a random 
noise. The effective Hamiltonian is obtained  by the 
standard Wigner--Weisskopf reduction~\cite{WIGNER,LARSSON,LOWDIN} by
partition.  
In this method the total Hilbert space 
is divided into two subspaces and the Hamiltonian is integrated over the
elements of one of them, eventually mapping the eigenvalue problem of
high dimension onto a lower one.
Models of such type are widely used in
e.g. condensed phase dynamics \cite{KARPLUS,BRICKMANN} where one is
typically interested in the detailed dynamics of only a small part of
the overall system, so one partitions the total system into the
subsystem of interest and a bath:
\eq
H=H_{sys} + H_{bath} + V
\eqx
Here $H_{sys}$ and $H_{bath}$ are the Hamiltonians of the isolated
subsystem and the bath; $V$ describes their interaction. The appropriate
scheme for partitioning is not always obvious~\cite{WEISS}; successful
choice minimizes the coupling $V$ so that its effects can be studied
perturbatively. After the separation, the state of the system is
described by its reduced density matrix which is obtained from the
complete system-plus-bath density by taking the trace over the bath
variables. When the relaxation of the bath is rapid compared to the
evolution of the system, the bath can be assumed to be equilibrated at
all times. Treating the interaction of the system and bath
perturbatively to the second order, the evolution of the reduced density
matrix can be then recasted in the form of the dissipative
Liouville-von-Neumann equation~\cite{REDFIELD}.

\noindent Similar ideology is used {\sl e.g.} to study the electron
or wave passage in disordered medium. In the field of chemical physics
the condensed phase reactions, 
such as electron and proton transfers  are of primary interest and
are commonly investigated in terms of a dissipative two-level system,
often referred to as spin-boson problem~\cite{KARPLUS,LEGGETT}.

\noindent We solve the analogue of this problem by investigating
analytically an effective two-level system coupled
to a Gaussian dissipative term. We demonstrate the phase transition in
such a system corresponding to the structural change in the average
spectral distribution of the eigenvalues. We present also the results of
the numerical simulations, confirming our analytical results. 
 Finally, section~5 outlines the possible applications and
generalizations of the considered model.

\section{Addition laws for random ensembles.}

The generic problem of finding the Green's function of a sum of
two independent hermitean matrices (here Hamiltonians) of dimension $N$
\eq
\phi_1+\phi_2
\label{sys12}
\eqx
has been considered in the seminal work by Voiculescu~\cite{VOICULESCU},
 and recently popularized by Zee~\cite{ZEE}. 

\noindent The main idea of Voiculescu was to find an elegant way of 
``linearizing'' the process of determining the average eigenvalue
distribution for the convolution of non-commuting operators, hence
finding an analogue of the logarithm of Fourier transformation for the
usual convolutions. As an example, let us consider two hermitean
operators in matrix representation, (\ref{sys12}), 
populated from the statistical ensemble  $P(\phi_1,\phi_2)$, provided
that in the large $N$ limit $P(\phi_1,\phi_2)$ factorizes into
$P_1(\phi_1)\cdot P_2(\phi_2)$. 
Then the resolvent
\begin{eqnarray}
G(z):=\frac{1}{N} \int d\phi_1 d\phi_2 P(\phi_1,\phi_2) 
	{\rm Tr}\frac{1}{z-\phi_1-\phi_2}   
\label{res}
\end{eqnarray}
could be expressed (at least {\em a priori}) knowing only the individual
resolvents
\begin{eqnarray}
G_i(z)=\frac{1}{N} \int d\phi_i \,P_i(\phi_i)\ {\rm Tr}
	\frac{1}{z-\phi_i} \qquad {\rm i=1,2} \,.
\label{res12}
\end{eqnarray}

\noindent The physical interpretation of the addition law comes from the
observation that 
\ben
G(z)=\frac{1}{N} \ {\rm Tr \,} \left< \frac{1}{z-\phi}
	\right>\equiv\frac{1}{z-\Sigma(z)}
\label{resolvent}
\een
where
\ben
\left< \ldots \right> = \int  P(\phi)  \ldots [d \phi]
\label{weight}
\een
and $\Sigma(z)$ is nothing but the self-energy. Through his diagrammatic
analysis, Zee \cite{ZEE} introduced the "Blue's function" that is just
the functional inverse of the resolvent
\eq
B(G(z))=z
\label{inverse}
\eqx
and satisfies the additivity law
\eq
B_{1+2}(z)=B_1(z)+B_2(z)-\f{1}{z}
\label{sumblue}
\eqx

\noindent  The operational algorithm is extremely simple when using the
Blue's function approach: 
First, having both ensembles, construct the two resolvents $G_1$ and $G_2$.
Second, invert them functionally, resulting in $B_1,B_2$. Third, add
them using the ``addition law'' (\ref{sumblue}). Fourth, invert functionally
the $B_{1+2}$, getting the Green function for the sum $G_{1+2}\equiv G$.
Fifth, get the average spectral distribution from G, using the standard
construction
\ben 
\varrho(\ee)= 
	-\frac{1}{\pi} \lim_{\lambda \rightarrow 0} {\rm
	Im\,}G(z)|_{z=\ee+ i\lambda}. 
\label{standard}
\een
The generalization for the {\bf nonhermitean case}  
amounts to considering the isospin Green functions
which now form a 2 by 2 matrix in isospin space, indexed by 
$z$ and $\bar{z}$. The generalized Blue's
function~\cite{JANIK}
 is now a matrix valued function of a $2\times 2$ matrix variable
defined by
\eq
\BB(\GG)=\ZZ=\arr{z}{0}{0}{\zb}
\eqx
The Green's function is 
found from the matrix version of (\ref{sumblue}):
\eq
\arr{z}{0}{0}{\zb}=\BB_1(\GG)+\BB_2(\GG)-\f{1}{\GG}
\label{e.add}
\eqx
In general, two kinds of the solutions to this matrix equations are 
possible. The first, when $\GG$ is diagonal, reproduces the 
hermitean construction  modulo the anti-analytical copy in $\bar{z}$, 
totally decoupled from the $z$-solution. Therefore this one reproduces 
the hermitean (or antihermitean case) only. The relevant solution
is the one when the off-diagonal element of the Green function is non-zero. 
In this case, the resolvent is nonanalytic, being the function of both 
variables $z,\zb$.  The support of the eigenvalues forms the
two-dimensional ``islands'' of nonanalyticity, (opposite to the
hermitean case, when only cuts constitute the support of real eigenvalues)
and the density distribution could be read-out from the 
2-dimensional ``Gauss law''~\cite{SOMMERS}
\ben
{\rm div E} = 4\pi \varrho(z,\zb)
\label{gauss}
\een
In this two-dimensional electrostatics divergence means differentiation 
with respect to $\zb$, and electric field has two components,
$E_x= 2 {\rm Re \,} \GG_{zz},\,E_y= - 2 {\rm Im\,} \GG_{zz}$.

\noindent Note that the non-analytical solution is a consequence of the 
``spontaneous breakdown of analyticity'' \cite{JANIK}. Naively, the
resolvent (\ref{resolvent})  is the function  
of $z$ variable only. But the true ground state is not respecting 
holomorphic separability into $z$ and $\bar{z}$ copies, resulting in 
the spontaneous breakdown of this ``symmetry'', by forming  the mixed  
``condensates'' $\GG_{z \bar{z}}$ and therefore 
breaking the analyticity (non-analyticity) of the solution\footnote{In
other words, the free energy associated with the non-analytical solution
lies below the one related to the analytical resolvent in the domain of
 the complex plane.}.
Below we demonstrate this scenario in the case of Wigner--Weisskopf
 reduction.

\section{Dissipative two-level system.}

We consider the  Hamiltonian with the Hilbert space spanned by $N$
discrete states $\ket{k}$ and $M$ continua $\ket{\chi_n(E)}$:
\ben
\HH=\sum^N_{k=1} E_k \ket{\psi_k}\bra{\psi_k} + \sum^M_{n=1} 
	\int dE E \ket{\chi_n}\bra{\chi_n} 
 \nonumber \\
+ {\gamma}^{1/2}\sum^M_{n=1} \sum^N_{k=1}\int dE
	(V^n_k(E) \ket{\psi_k}\bra{\chi_n} + h.c.) \,.
\een
The bound continuum coupling is characterized by the coupling constant
$\gamma$ and the energy dependent matrix $V^n_k(E)$. Note that in this
model there is no direct coupling (interactions) between continuum
channels $n, n'$. After eliminating the continua 
$\ket{\chi_n(E)}$, the effective Hamiltonian $H_{eff}$ takes the form
\ben
H_{eff} = \HH-i\gamma VV^{\dagger}\,,
\label{beginning}
\een
where $\HH$ is $N\times N$ deterministic Hamiltonian with the bound-state
energies $\ee_1,\ldots,\ee_N$ and the dissipative part is obtained via
so-called
Markovian\footnote{The Markovian property assumes exponential relaxation
of the response of the continuum when it acts on the test system and
rephrased in thermodynamic terms, would correspond to the high
temperature Debye relaxation bath. In the above model, Markovian
approximation should be understood as the condition for almost
instantaneous decay of the kernel in the evolution (Schr\"odinger)
equation satisfied by the bound-space state vectors $\ket{\psi_k}$. The
condition is met if the energy dependence of the matrix elements
$V^n_k(E)$ is negligible \cite{BLOCH,REDFIELD,MAHAUX}.} limit.
The operators $V^n_k =\bra{\psi_k}V\ket{\chi_n}$ stand for the coupling
of the bound-state eigenvectors $\psi_k$ ($k=1,\ldots, N)$ 
corresponding to eigenvalues $\ee_1,\ldots,\ee_N$
with the continuum states $\chi_n$ ($n=1,\ldots, M$) and account for the
depletion of the bound state subspace due to the leakage of the
probability mass into the continua. The eigenvalues $\ee=x-iy$ of the
effective Hamiltonian $H_{eff}$ are in general complex, their imaginary
parts must be positive for damped systems and give the lifetimes of the
corresponding eigenvalues proportional to $1/y$.

\noindent The model Hamiltonian of that type (\ref{beginning}) is widely
used in quantum chaotic scattering problems \cite{MAHAUX} but has been
also introduced to study unimolecular dissociation of selectively
excited polyatomic molecules \cite{LEVINE,PESKIN}. For most
dissociating molecules above the tunneling regime (when the excitation
energy is well above the dissociation threshold) the coupling between
the molecular complex and the continuum is strong and the density of
strongly mixed molecular states is large, so that the unimolecular decay
is characterized by overlapping resonance states.
   
\noindent We will solve the system in the regime  when both $M$ and $N$
are large, but  $m=M/N$ is fixed (which implies strongly overlapping
resonances). To simulate the effect of the "noise" represented by
statistical properties of the operators $V, V^{\dagger}$ we will assume
that the N by M matrices $V$ are random. For simplicity we will choose
randomness as Gaussian \footnote{By  use of the central limit theorem,
Gaussianity is a legitimate choice for the infinite normalized sum of
independent fluctuating contributions. It can be shown also that the
distribution for normalized sum of matrices $\phi_{sum}\equiv
\frac{1}{\sqrt{k}} \Sigma_i^k \phi_i$, each sampled from a distribution
$P_i (\phi_i)$ follows {\cite{ZEE,BREZIN}} the Gauss theorem of
probability. Using measures different from the Gaussian one do not
change the addition law discussed in this section.}, thus working within
the Gaussian unitary ensemble (GUE). Furthermore, we will investigate a
two-level system, with deterministic eigenvalues
$\ee_1=\ldots=\ee_{N/2}\equiv\ee$ and 
$\ee_{N/2}=\ldots=\ee_N\equiv -\ee$. 

\noindent To apply the addition law (\ref{e.add}), we need only the
functional form of the corresponding Blue's functions. For the
deterministic part, the Green function is by definition 
\ben
\GG_D(\ZZ)=\frac{1}{2} \left(\frac{1}{\ZZ-\ee} +  
	\frac{1}{\ZZ+\ee}\right).
\label{greendet}
\een
where $z$ has been replaced by the matrix ${\cal Z}={\rm diag} (z, \bar{z}))$.
Therefore the relevant Blue's function is given by the equation
\ben
\ZZ=\frac{1}{2}\left((\BB_D(\ZZ) -\ee)^{-1} +  (\BB_D(\ZZ) +\ee)^{-1}\right)
	\,.
\label{what}
\een
The Blue's function for the noise term $i\gamma VV^{\dagger}$ can be
determined by use of diagrammatic techniques and has been first derived
in \cite{JANIK}:
\ben
\BB_{VV^\dagger}= m(1-\Gamma\ZZ)^{-1}\Gamma+\frac{1}{\ZZ}.
\label{bluenoise}
\een
where $\Gamma$ is the coupling matrix
\ben
\Gamma=\arr{-i\gamma}{0}{0}{i\gamma} \,.
\label{GA}
\een

\noindent Note that all above equations have been written in the matrix
form, in agreement with the discussion of the previous section.

\noindent We can further use the generalized addition formula
(\ref{e.add}) for the total system, resulting in 
\ben
\BB(\ZZ)=\BB_D(\ZZ)+m(1-\Gamma\ZZ)^{-1}\Gamma \,.
\label{ff}
\een
Inverting that relation ({\it i.e.} replacing $\ZZ$ by $\GG$), we get
\ben
\ZZ=\BB_D(\GG)+m(1-\Gamma\GG)^{-1}\Gamma \,.
\label{eq.1}
\een
By similar replacement in equation (\ref{what}) we get
\ben
\GG=\frac{1}{2}\left((\BB_D(\GG) -\ee)^{-1} +  (\BB_D(\GG) +\ee)^{-1}\right)
	\,.
\label{eq.2}
\een
Calculating $\BB_D(\GG)$ from (\ref{eq.1}) and plugging it into 
(\ref{eq.2}) we arrive at the final equation for the Green function
\ben
\GG=\frac{1}{2}\left((\ZZ-m(1-\Gamma\GG)^{-1}\Gamma -\ee)^{-1} +
  (\ZZ-m(1-\Gamma\GG)^{-1}\Gamma +\ee)^{-1}\right) \,.
\label{final}
\een
The matrix equation~(\ref{final}) is a central formula to the paper from
which all 
further discussion of spectral properties of the system will be derived.
We solve this equation using the explicit parameterization of the 
matrix $\GG$ with the unknown entries
\ben
\GG=\arr{a}{b}{d}{c} \,.
\label{GGm}
\een

\noindent As already mentioned in the previous section, the solutions
belong to two classes: 
\vskip 1cm
\noindent $\bullet$\,\,\,{Holomorphic (analytical) case.}

\noindent This yields a trivial solution to the problem. In this case,
we seek solutions with an imposed condition $b=d=0$.
The above
matrix equation splits then into two mirror copies of algebraic equations for
$z$ and $\bar{z}$, respectively. The equation for $a=\GG_{zz}$ is of
Cardano type 
\ben
 a^3({\ee}^2\delta^2 - \delta^2 z^2)+ a^2(\delta^2 z+2\delta z^2-2m\delta^2 z-
2\ee^2\delta) \nonumber \\
+ a(m\delta^2+2m\delta z-m^2\delta^2-2\delta z-z^2+\ee^2) +
 z-m\delta=0  \nonumber \\
 \delta = -i\gamma
\een
with the Blue's function 
\ben
B(z)=\frac{m\delta}{1-\delta z} + \frac{1}{2 z} + \biggl(\frac{1}{4z^2}
	+\ee^2\biggr)^{1/2} . 
\een

\noindent Solution to those equations constitute nothing but a complement to
the dashed domains of density $\varrho(z,\bar{z})$ pictured in Fig.~1. 
We reject it as a non-physical one. The proper solution to our problem
follows a non-analytical case which we are going to discuss now.
%

\begin{figure}[htbp]
\centerline{\epsfysize=8truecm \epsfbox{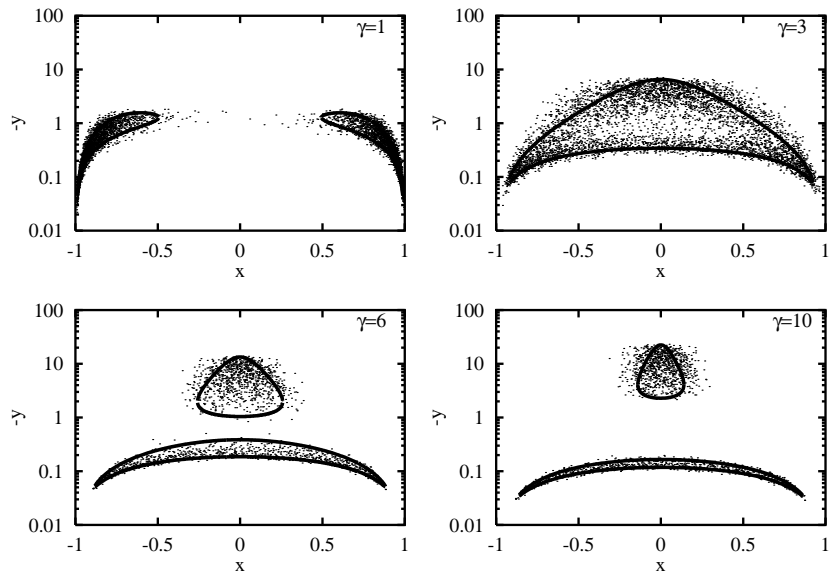}}
\caption{The support of the eigenvalues of the two-state system for
different coupling at fixed filling ratio $m=0.25$. The dots correspond to
the numerical simulation of the Hamiltonian~(\protect\ref{beginning}),
while the solid lines come from the analytical
result~(\protect\ref{bndry}). Numerical simulations have been performed for
100 matrices of size $N=100$ at $\gamma =1, \gamma=3$ (the upper row of
drawings), and for 30 matrices with $N=400$ at $\gamma=6$ and
$\gamma=10$ (the lower row). The density of the dots is related to the 
the spectral density~(\protect\ref{spdens}).}
\label{fig.cudna}
\end{figure}

\vskip 0.5cm
\noindent $\bullet$\,\,\,{Non-holomorphic (non-analytical) case.}

\noindent This case corresponds to a situation when,  despite the
resolvent naively is the function of the $z$ variable only, the support
of the spectrum depends on both $z$ and $\bar{z}$, forming therefore
two-dimensional islands (in contrast to one-dimensional cuts of the
analytical solution) in the complex plane. The solution of the
matrix equation now is given by
\ben
a&=& \frac{x}{\gamma y} - \frac{x}{2(x^2-\ee^2)} +
\frac {i}{2}\biggl(\frac {1}{\gamma}+ {\frac {1}{\gamma}}{\frac{x^2-{\ee}^2}
{y^2}} + \frac {m-1}{y}\biggr)\nonumber \\
c&=&\bar{a}
\een
and $b=d$, where $b$ satisfies the equation
\ben
b^2=\frac{m}{\gamma y}+|a|^2 + \frac{1}{\gamma^2} + \frac{i}{\gamma}(a -
	\bar{a}) \,.
\een
The spectral density obtained from Gauss law (\ref{gauss}) reads
\ben
\varrho=\frac{1}{4\pi}\biggl [\frac{2}{\gamma
y}\biggl(1+\frac{x^2-\ee^2}{y^2}\biggr)+\frac{m-1}{y^2}
+ \frac{x^2+\ee^2}{(x^2-\ee^2)^2}\biggr]
\label{spdens}
\een

\noindent The shape of the envelopes ({\sl i.e.} boundaries of 2-dimensional
regions of spectral densities in the complex plane) is given by the condition
$b=\GG_{z\bar{z}}=0$ which is equivalent to
\ben
\frac{x^2-\ee^2-\gamma y}{\gamma^2 y^2} = \frac{x^2}{4}
	\biggl(\frac{2}{\gamma y} - \frac{1}{x^2 -\ee^2}\biggr)^2
	+\frac{1}{4} \biggl[\frac{1}{\gamma} 
	\biggl(1+\frac{x^2-\ee^2}{y^2}\biggr)+
	\frac{m-1}{y}\biggr]^2 \,.
\label{bndry}
\een
The spectrum undergoes two various phase transitions 
({\sl cf.}~Fig.~\ref{fig.cudna}). For very low 
coupling, the energy islands do not connect the two states and only
after reaching the critical value of the coupling $\gamma$,
a single island of the eigenvalues is being created (see the upper row
of drawings in Fig.~\ref{fig.cudna}).
We note, that the global energy scale $\epsilon$ may be eliminated
in~(\ref{bndry}) by introducing the dimensionless variables $x'=x/\epsilon$,
$y'=y/\epsilon$ and $\gamma'=\gamma/\epsilon$.

\begin{figure}[htbp]
\centerline{\epsfysize=5truecm \epsfbox{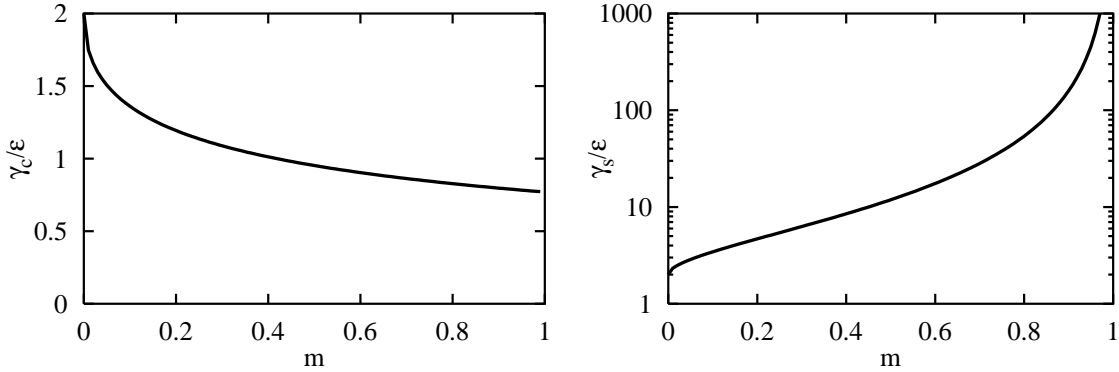}}
\caption{The dimensionless critical couplings of the two-state
model as functions of the filling ratio $m$. The left figure presents
the critical coupling  $\gamma_c$ at which the corridor between the two 
state is built up, the right figure displays the critical coupling
$\gamma_s$ at which the cloud of resonance states splits.}
\label{f:gcrit}
\end{figure}

\noindent The second critical point (see the lower row of drawings 
in~Fig.~\ref{fig.cudna})
 corresponds to the situation when
one island of non-analyticity splits into two, a low lying broad island and a
narrow island of short-living states. The critical values of the
coupling constant can be determined from the study of the
discriminant of the boundary equation (\ref{bndry})
at $x=0$. The zero value of discriminant appears if distinct roots of
the equation coincide, thus it is suggestive of occurrence of phase
transitions discussed in the text. The condition is described by a
second order equation in $\gamma^2$,
\ben
  (1-m)^3 \gamma_*^4+(m^2-20m-8) \epsilon^2 \gamma_*^2 +16 \epsilon^4 = 0\,.
\label{disc}
\een
The solution $\gamma_*$ scales with the energy scale $\epsilon$, and the
critical dimensionless coupling $\gamma_*/\epsilon$ is
universal. Figure~\ref{f:gcrit} shows the two critical couplings as the
function of the filling ratio $m$. The lower value $\gamma_c$, corresponding
to the connection of the two energy states, starts at value 2 and goes to
$4/3\sqrt{3}$ as $m\to 1$. The other branch of critical coupling
parameter $\gamma_s$, describing the
splitting of the broad low lying island and the narrow, short-living
region, starts at the same value as $\gamma_c$ and goes to infinity as
$m\to 1$. For small values of $m$ ($m\to 0$), appearance of bridging
states between the two "deterministic" states results in the immediate
splitting of the cloud of the resonance states. Distribution of points
within the clouds follows the ratio $(1-m) : m$ between the low
lying broad cloud and the short-living upper cloud, respectively. 

\noindent By integrating $\varrho(x,y)$ over the support of $x$ we
obtain the distribution of resonance widths in the model.
\begin{figure}[htbp]
\centerline{\epsfysize=8truecm \epsfbox{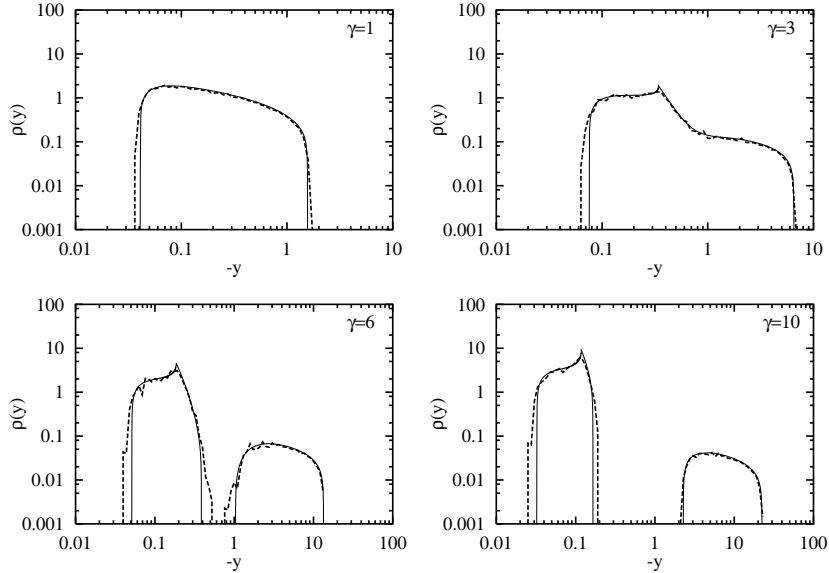}}
\caption{The integrated spectral density, $\varrho(y)$ describing the
density of the resonance widths in the system governed by
$H_{eff}$. Solid line stands for the analytical prediction, the dashed
line represents numerical estimate obtained by use of an ensembles of 30
random matrices with size $N=400$.}
\label{f:integ}
\end{figure}

\begin{figure}[htbp]
\centerline{\epsfysize=8truecm \epsfbox{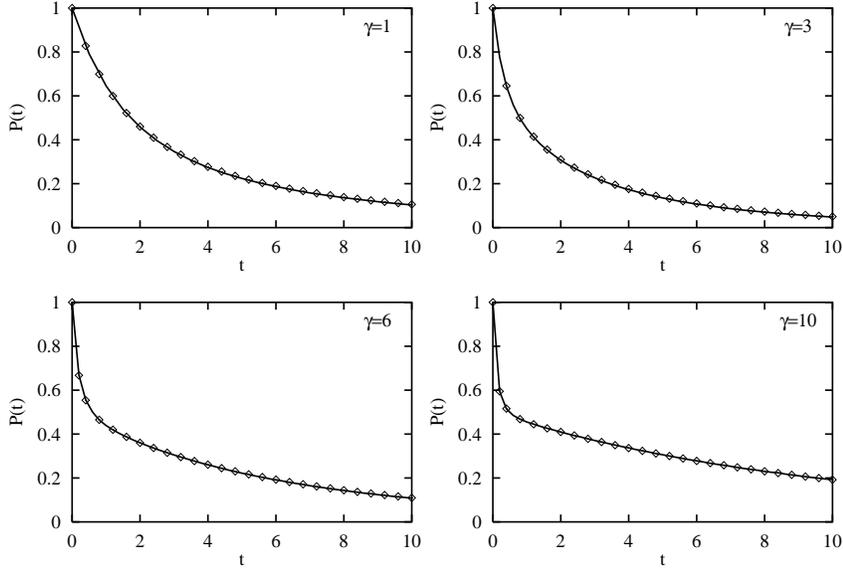}}
\caption{The decay of the survival probability (\ref{prob}) for various 
values of the coupling constant $\gamma$. Nonexponential decay becomes
transparent for increasing values of $\gamma$.The solid line is the
Laplace transform of the analytical width distribution (\ref{lap}). The
dots correspond to the Laplace transform of the density obtained
numerically with a set of 30 random matrices of size $N=400$.}
\label{f:lapl}
\end{figure}

\noindent Using the symmetry
of the spectra we arrive at
\ben
\varrho(y) &=& \int_{\makebox[0mm]{\raisebox{-2mm}{support}}} dx\
	\varrho(x,y) = R(x_u,y)-R(x_l,y) \,, \\
R(x,y) &=& \frac{1}{2\pi} \left\{ \frac{2x}{\gamma y} + 
	\frac{2x}{\gamma y^3} \left(\frac{x^2}{3} - \epsilon^2\right)
	-\frac{(1-m)x}{y^2}+\frac{x}{\epsilon^2-x^2} \right\} \,,
	\nonumber
\label{lap}
\een
where $x_u$ and $x_l$ are the two positive end points of the support
(for a continuous support $x_l=0$). The resulting density
can be used to determine average decay rate in the system, and is shown
in Fig.~\ref{f:integ}.
\noindent The bound state component $\ket{\psi}$ evolves in time according to
\ben
\ket{\psi(t)}=e^{-iH_{eff}t/\hbar}\ket{\psi(0)}
\een
and its effective decay rate can be defined as the logarithm derivative
of its norm
\ben
k_{eff}=-\frac{d}{dt} \ln \la\psi(t)|\psi(t)\ra \,.
\een
If the initial state has nonzero overlap with only one eigenstate of
$H_{eff}$, the above definition simplifies to $k_{eff}=2y_i/\hbar$. In
general case when resonances overlap, the average survival probability
over the ensemble is given by
\ben
P(t)=\int^{\infty}_0\varrho (y) \exp(-yt) dy
\label{prob}
\een
and is related to the distribution of kinetic rates $\varrho(y)$  by the
Laplace transform. The result is displayed in~Fig.~\ref{f:lapl} showing
the decay with some average decay rate for low couplings and the
appearance of roughly two distinct decay rates for large couplings. 
The time scale is of the order of a femtosecond for $\ee=1eV$, and
scales linearly with energy. (Note, that within the paper the
mean-square dispersion of elements of $V$ matrix has been put equal 1.)

\begin{figure}[htbp]
\centerline{\epsfysize=8truecm \epsfbox{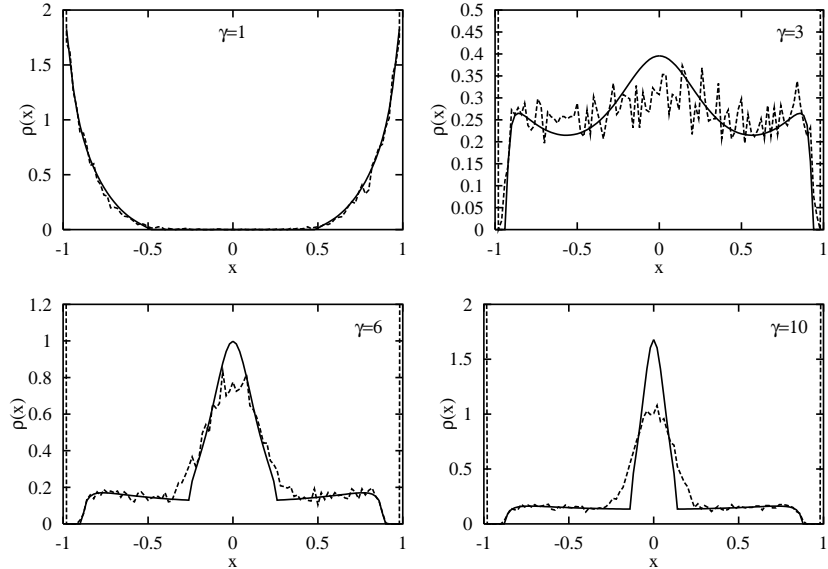}}
\caption{The integrated spectral density, $\varrho(x)$ describing the
total width of a given energy state.
Solid line stands for the analytical prediction, the dashed
line represents numerical estimate obtained by use of an ensembles of 30
random matrices of size $N=400$.}
\label{f:integy}
\end{figure}

\noindent By integrating the spectral density~(\ref{spdens}) along the width
coordinates one can obtain the total width assigned to a given
energy. The resulting density reads
\ben
\varrho(x) &=& \int_{\makebox[0mm]{\raisebox{-2mm}{support}}} dy\
	\varrho(x,y) = \sum_{i=1,2} R_x(x,y_{2i+1})-R_x(x,y_{2i+2}) \,, \\
R(x,y) &=& \frac{1}{4\pi} \left\{ \frac{\ee^2-x^2}{\gamma y^2} + 
	\frac{1-m}{y} + y\frac{\ee^2+x^2}{(\ee^2-x^2)^2} +
	2\frac{\mbox{ln}\,|y|}{g} \right\} \,,
	\nonumber
\een
with ($y_1$--$y_2$) and ($y_3$--$y_4$) the supports of the
integration. The inverse of that quantity gives information about the
decay time of the state, labeled with the given energy. The distribution
of the total width as the function of the energy at different couplings
is presented in Fig.~\ref{f:integy}.
\noindent The situation depicted in Fig.~\ref{f:integy} strongly resembles
 ``noise induced transition'' when coupling to a fluctuating source
induces new order in the system. For low values of $\gamma$, distribution
$\varrho(x)$ has maxima around eigenvalues of the deterministic Hamiltonian;
that bimodality of $\varrho(x)$ changes with increasing value of $\gamma$
which introduces new maximum around $x=0$, strongly populated for high
values of the coupling constant.

\section{Collectivization of the spectra}

At large coupling constant the spectrum shows the structural change. Two
groups of distinct eigenvalues appear. The first, saturating the most of
the total width corresponds to the direct resonances. The second group
corresponds to long-living states and appears only, if the coupling is
sufficiently strong. Such a "reorganization" of the spectrum is usually
observed~\cite{SOB,REMACLE,SOKOLOV,FYODOROV} in systems with $N$
degenerate bound states coupled to $M$ continua. The system develops
then $M$ decaying states and $N-M$ stable states whose population
becomes trapped. The similar phenomenon of formation of a short-lived
coherently decaying
state in a system with almost degenerate levels is known in quantum
optics (Dicke superradiance \cite{DICKE}), nuclear physics (giant dipole
resonances, collective isobar-hole states in scattering at intermediate
energies \cite{ZELEVINSKY}) and dissipative spin systems
\cite{GROBE}. An interesting aspect of the phenomenon brings also a
semiclassical description of collectivization discussed in papers of
{\it Gaspard and Rice} \cite{RICE}. Multiple hoppings lead to very long
trajectories (the delay time is large) saturating the Gutzwiller formula
and causing breakdown of the ergodicity. 
The physical mechanism responsible for the appearance of the above
mentioned structural change in the spectra is caused by the presence of
zero modes in the matrix $VV^{\dagger}$. If the coupling is
strong, the zero modes are becoming delocalized, corresponding to the
appearance of the coherent, collective state. In other words, for increasing
coupling $\gamma$, the probability of hopping from one zero mode to
another (or effective "overlap" between the zero-mode states) is
increasing, and the long strings due to multiple hopping can
appear. This mechanism is somewhat analogous to the Mott's
conductor-insulator phase transition, when the conductance appears due
to non-zero probability of the hopping of the electron between the
various Fermi levels, so that the electrons are no longer
localized. Another physical example corresponds to the picture
of spontaneous breakdown of the chiral symmetry in quantum
chromodynamics \cite{NOWAK}, when chiral condensate appears as a
coherent effect of the delocalization of the zero modes of gluonic
excitations. 

\section{Conclusions}

\noindent Our analysis of the two-level system coupled to a random noise shows
the advantage of using the Blue's function approach \cite{ZEE} to study
spectral properties of dynamical quantum systems. By use of the
generalized Blue's function~\cite{JANIK}, we are able to construct
easily the Green function relevant for the system and from the last one
we can read off spectral properties of the effective Hamiltonian.

\noindent Characteristic  features of the distribution function of the
eigenvalues related to the effective Hamiltonian depend strongly on the
intensity of the coupling between the deterministic and random parts of
the Hamiltonian. At low values of coupling constant, we observe some
dispersion of eigenvalues density around deterministic energies of the
two-level system. Beyond the first critical value of $\gamma=\gamma_c$,
a common region connecting "dispersed deterministic eigenvalues" is
created and builds up with further increasing $\gamma$. At very high
values of coupling, dissipativity entering the system through the term
$VV^{\dagger}$ leads to reorganization of eigenvalues. We observe
"collectivization" of widths to a bridging, narrow region which is
populated by low energy states. Its presence suggests then formation of
long-living states, well separated from the cloud of resonances
responsible for a rapid decay.
\noindent  Despite its simplicity, the model presented here can cover
vast number of various applications. We believe that one of direct use
of the analysis is modeling unimolecular decay processes in polyatomic
molecules and electron tunneling processes in biological media. Studies
of an effective tunneling matrix in biomolecules require knowledge of
the structure of off-diagonal elements of the generalized Green function
discussed above. Analysis of the kinetic rate in such models is based on
the form of two-particle (two-time) correlation function which, for the
model Hamiltonian of "deterministic plus random" dynamics, will be
discussed in the forthcoming paper.

\vskip 1.2cm
{\bf \noindent  Acknowledgments } \\ 
\vskip .3cm
\noindent
The authors acknowledge enlightening and instructive discussions with
R.~Janik, M.~A.~Nowak and I. Zahed on diagrammatic approach in the
theory of random
matrices. E.G-N. thanks U. Schmitt for many conversations on charge transfer
models in condensed media and S. Kast for bringing to our attention relevant
papers \cite{LEVINE,PESKIN}. 
G.P. acknowledges a partial support by the Hungarian Research Foundation
OTKA.
\vglue 2.5cm


\end{document}